\documentclass[twocolumn]{jpsj2}    %
\def \braket<#1>{\langle{#1}\rangle}
\def \v...{\mbox{\scriptsize $\vdots$}}
\def \c...{\mbox{\small $\cdots$}}
\newcommand  {\img}   {{\rm{i}}}
\newcommand  {\bhline}{\noalign{\hrule height 1.0pt}}
\renewcommand{\(}     {\left(}
\renewcommand{\)}     {\right)}
\renewcommand{\_}[1]  {_{\rm #1}}
\renewcommand{\^}[1]  {^{\rm #1}}
\newcommand  {\eqn}[1]{(\ref{eqn:#1})}

\title{Point-Contact Conductance in Asymmetric Chalker-Coddington Network Model}

\author{
 Koji \textsc{Kobayashi}\thanks{E-mail address: k-koji@sophia.ac.jp},
 Tomi \textsc{Ohtsuki}, and
 Keith \textsc{Slevin}$^{1}$
}

\inst{
 $\ \ $Department of Physics, Sophia University, Tokyo 102-8554, Japan \\
 $^{1}$Department of Physics, Osaka University, Osaka 560-0043, Japan
}

\abst{
 We study the transport properties of disordered two-dimensional electron systems with a perfectly conducting channel.
 We introduce an asymmetric Chalker-Coddington network model 
and numerically investigate the point-contact conductance.
 We find that the behavior of the conductance in this model is completely different from that in the symmetric model.
 Even in the limit of a large distance between the contacts, 
we find a broad distribution of conductance 
and a non-trivial power law dependence of the averaged conductance on the system width.
 Our results are applicable to systems such as zigzag graphene nano-ribbons 
where the numbers of left-going and right-going channels are different.
}

\kword{
 point-contact conductance,
 perfectly conducting channel,
 network model,
 Chalker-Coddington model,
 graphene
}

\begin{document}

\maketitle

\section{Introduction}
\begin{figure}[htb] 
         \begin{center}
          \begin{tabular}{@{\,}ccc}
             {\small (a) insulator} & {\small (b) criticality} & {\small (c) quantum Hall} \\
            \raisebox{2mm}{\small ($p \to 0$)} & \raisebox{2mm}{\small ($p=0.5$)} & \raisebox{2mm}{\small insulator ($p \!\to\! 1$)} \\
            \includegraphics[height=73pt,bb=0 0 340 367]{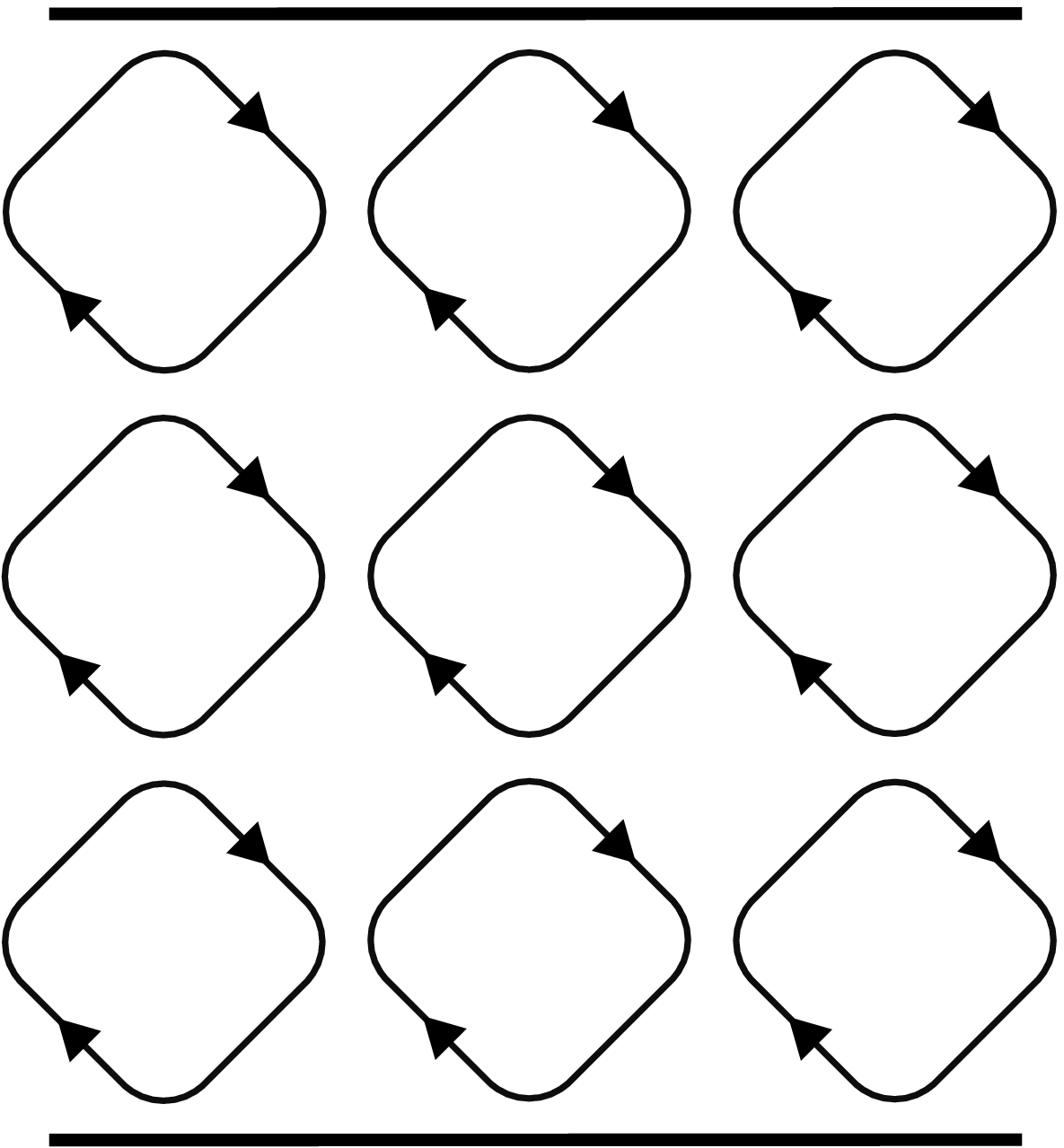} &
            \includegraphics[height=73pt,bb=0 0 341 367]{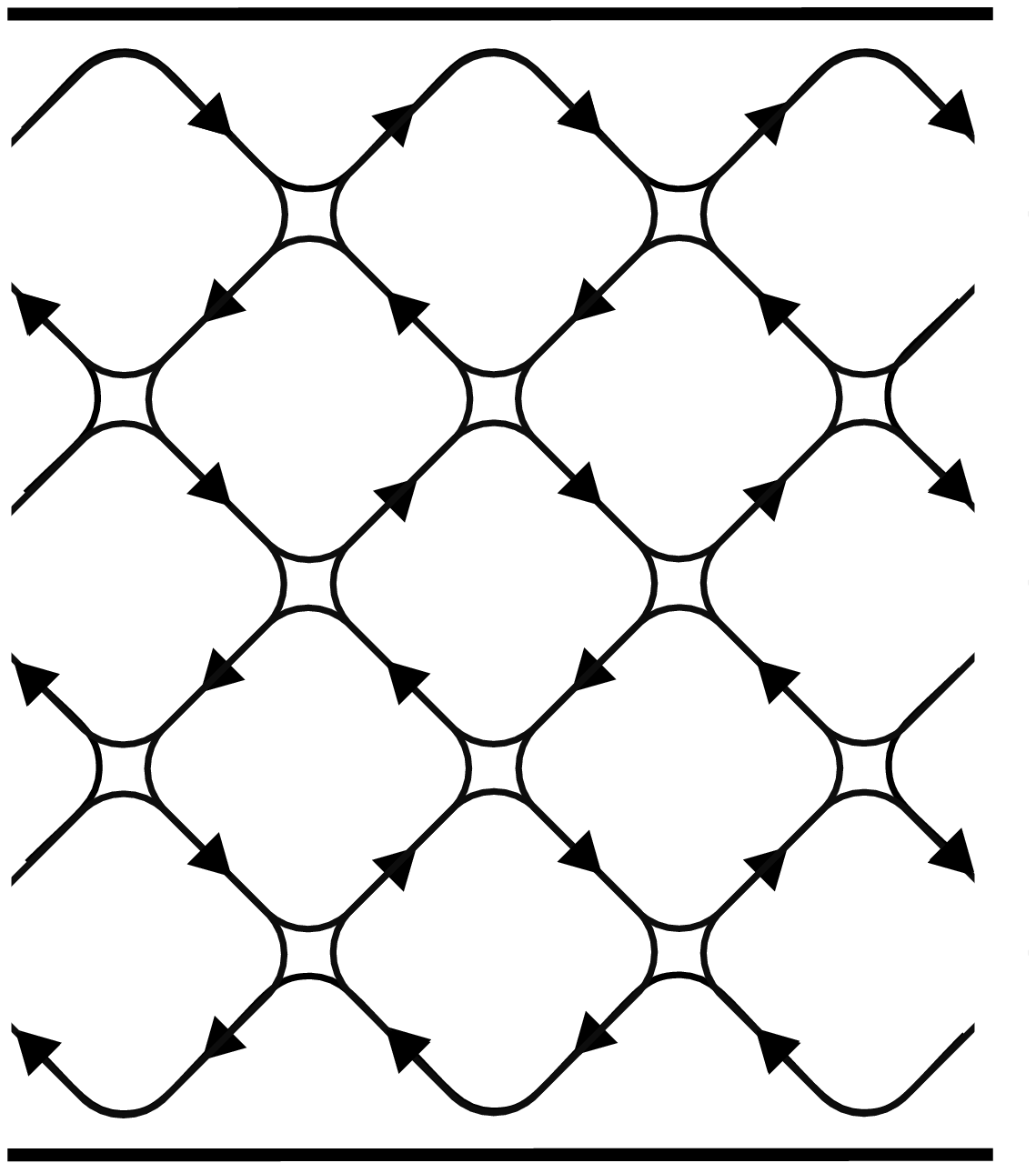} &
            \includegraphics[height=73pt,bb=0 0 314 367]{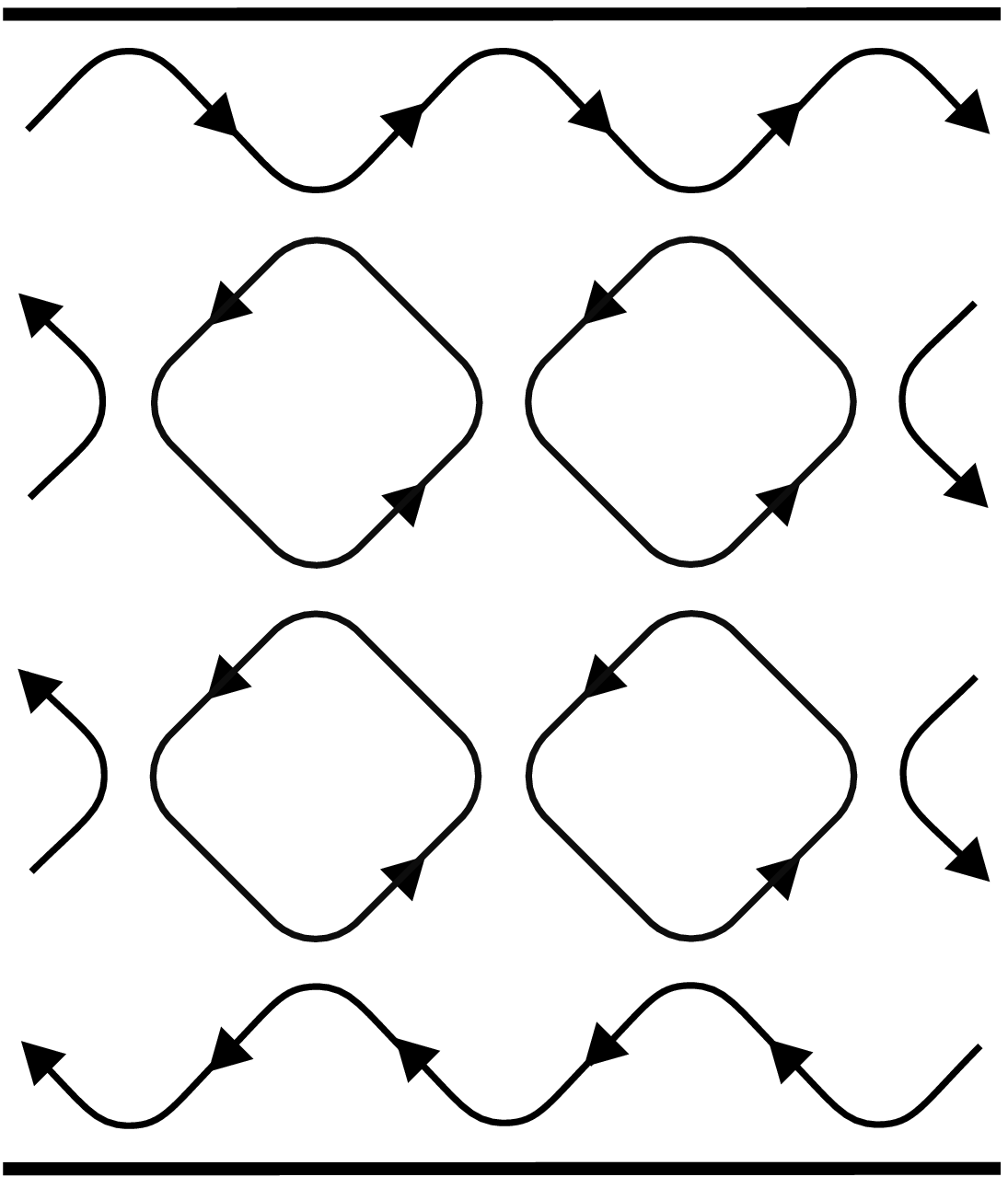} \\[2mm]
          \end{tabular}
         \end{center}\vspace{-5mm}
        \caption{ A schematic of the current flow in the CC model.
                 Current flows along the links: arrows indicate the direction of flow.
                 In the insulator (a), all current paths are closed and all states are localized,
                 while at criticality (b), all states are delocalized.
                 In the quantum Hall insulator (c), bulk states are localized, 
                 but the 
                 edge states carry current.}\vspace{-2mm}
        \label{fig:CCNWs}
        \end{figure}

 The critical behavior of the transport properties of two-dimensional electron systems under quantum Hall\cite{Klitzing_QHE} conditions 
has been investigated in various models.
 The Chalker-Coddington (CC) network model \cite{CC_NWM} (Fig.~\ref{fig:CCNWs}) 
is especially suited to the calculation of transport properties\cite{KOK} 
because current amplitudes are calculated directly.
 This is in contrast to the tight binding model where the wave functions must be calculated first.

\begin{figure}[htb] 
         \begin{center}
          \begin{tabular}{c}
            \includegraphics[width=65pt]{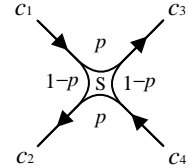} 
          \end{tabular}
         \end{center}\vspace{-4mm}
        \caption{Scattering at a node.
                 The current amplitude on link $i$ is $c_i$.
                 The incoming currents are scattered to the left with probability $p$ and 
                 to the right with probability $(1-p)$.}\vspace{-4mm}
        \label{fig:CCnode}
        \end{figure}

 The CC model consists of links corresponding to equipotential lines 
and nodes describing the scattering at saddle points of the random potential.
 Assuming the amplitudes of incoming and outgoing currents for a node to be $c_1, c_4$ and $c_2, c_3$, respectively 
(see Fig.~\ref{fig:CCnode}), 
scattering at a node is described by a $2\times 2$ unitary scattering matrix $\bf s$,
        \begin{equation} \label{eqn:s_1}
          \(\begin{array}{@{\,}c@{\,}}
           c_2  \\
           c_3
          \end{array}\)
          = {\bf s}
          \(\begin{array}{@{\,}c@{\,}}
           c_1  \\
           c_4
          \end{array}\),
        \end{equation}
        \begin{equation} \label{eqn:s_2}
         {\bf s} =
          \(\begin{array}{@{\,}c@{\ \ }c@{\,}}
            e^{\img\phi_2}\! & 0                  \\
            0                & e^{\img\phi_3}\! 
          \end{array}\)\!
          \(\begin{array}{@{\,}c@{\ \ }c@{\,}}
            \sqrt{1\!-\!p}  \!&  \sqrt{p}     \\
            \sqrt{p}        \!& -\sqrt{1\!-\!p}
          \end{array}\)\!
          \(\begin{array}{@{\,}c@{\ \ }c@{\,}}
            e^{\img\phi_1}\! & 0                  \\
            0                & e^{\img\phi_4}\! 
          \end{array}\).
        \end{equation}
 The effect of disorder is included in the phases $\phi_i$, 
which are independently and uniformly distributed between $0$ and $2\pi$.
 The scattering probability $p$ controls whether the system is 
in the insulating regime ($0 \le p < 0.5$),
at criticality ($p = 0.5$), or
in the quantum Hall insulating regime ($0.5 < p \le 1$).
%

\subsection{Asymmetric Chalker-Coddington network model}
\begin{figure}[htb] 
         \begin{center}
          \begin{tabular}{@{\,}ccc}
             {\small (a) asymmetric} & {\small (b) criticality} & {\small (c) asymmetric} \\
            \raisebox{2mm}{\small insulator ($p \!\to\! 0$)} & \raisebox{2mm}{\small ($p=0.5$)} & \raisebox{2mm}{\small insulator ($p \!\to\! 1$)} \\
            \includegraphics[height=61pt,bb=0 0 340 310]{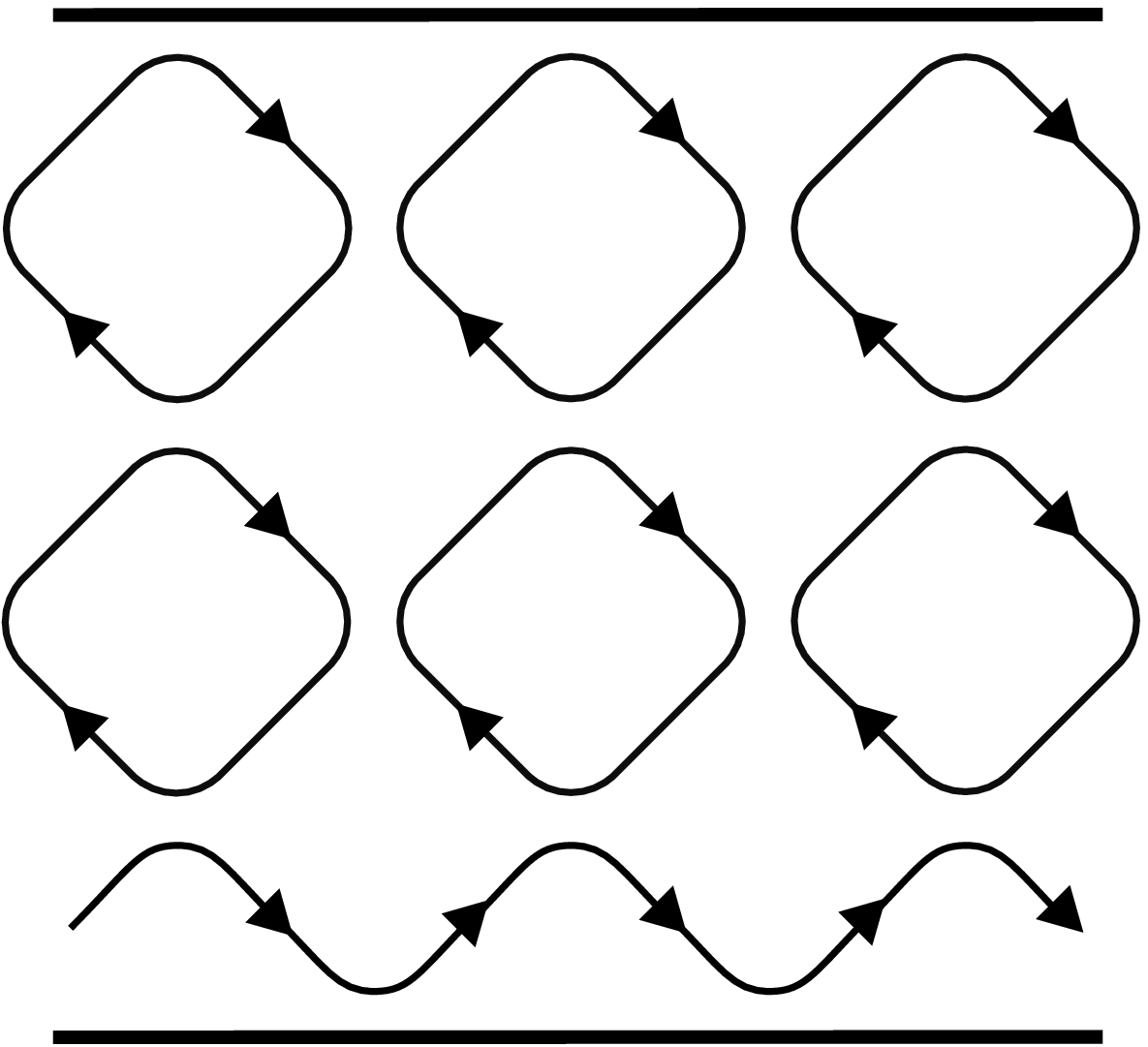}  &
            \includegraphics[height=61pt,bb=0 0 342 310]{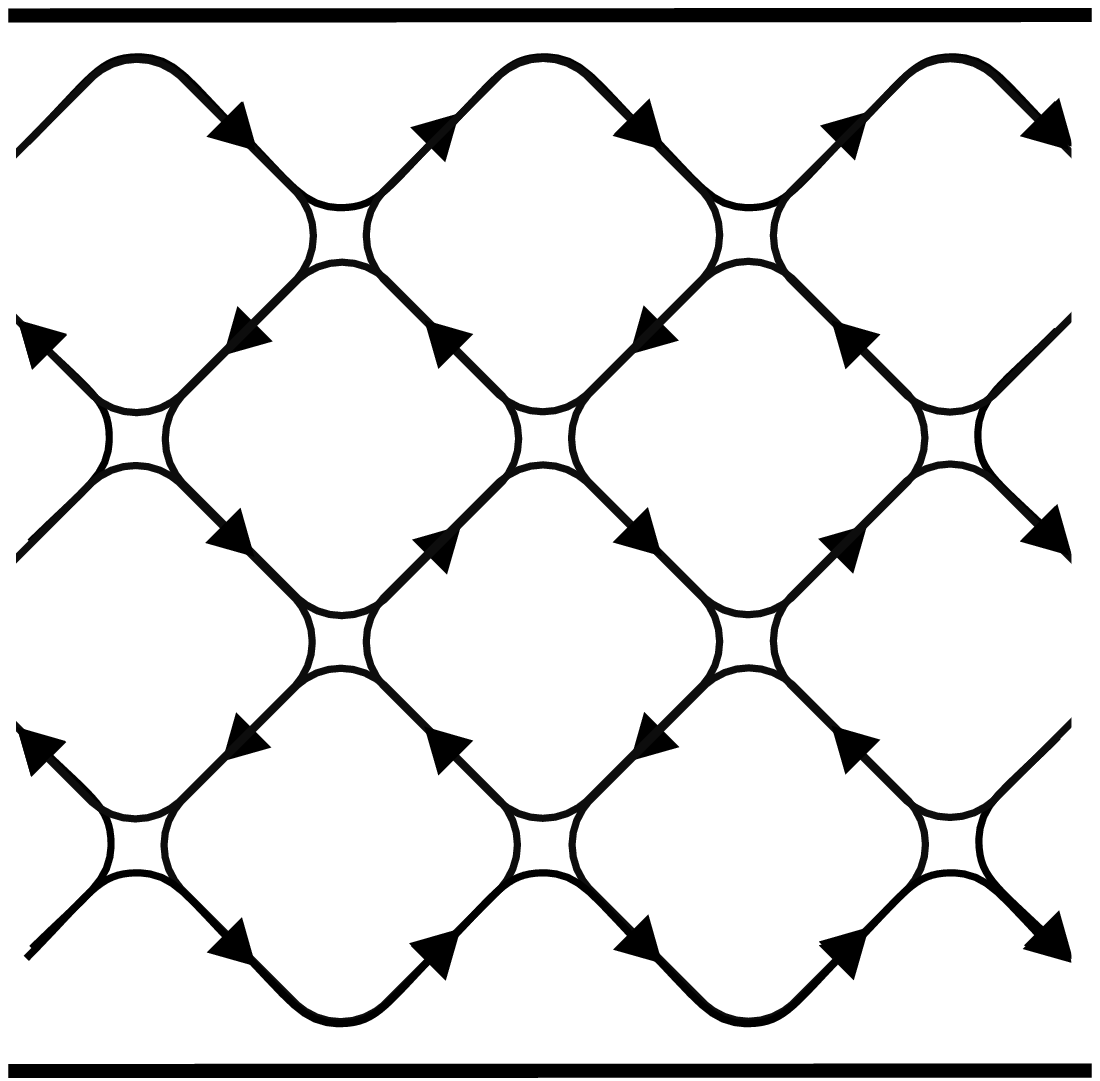}&
            \includegraphics[height=61pt,bb=0 0 314 310]{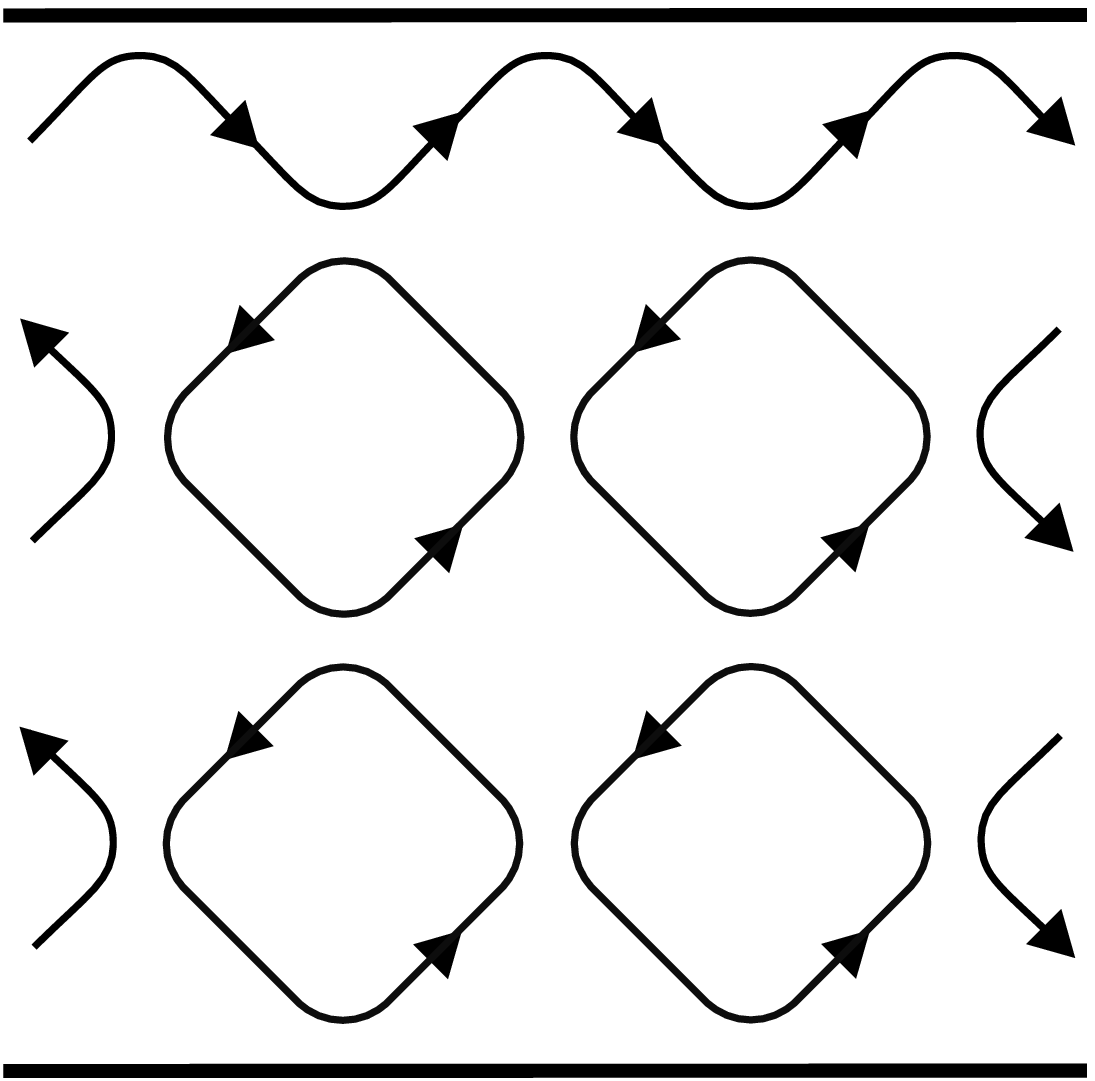}
          \end{tabular}
         \end{center}\vspace{-4mm}
        \caption{ A schematic of the current flow in the asymmetric CC model.
                 In the insulating phases (a) and (c), there is a conducting channel at one of the edges.}\vspace{-3mm}
        \label{fig:CCNWs_asym}
        \end{figure}

 In tight binding models, 
the numbers of the right-going and left-going channels are always the same.
 Under certain conditions, however, some of the channels decouple.
 One example is a graphene sheet with zigzag edges \cite{Wakabayashi_ZGNR}, 
where there are $l$, say, left-going and $l+1$ right-going channels near $ka=2\pi/3$, 
and $l+1$ left-going and $l$ right-going ones near $ka=-2\pi/3$, 
where $k$ is the wave number and $a$ the lattice constant.
 For long ranged scatterers, states near $2\pi/3$ and $-2\pi/3$ do not mix, 
and hence the numbers of right-going and left-going channels become, in effect, asymmetric.
 This asymmetric situation has been studied 
numerically for quantum railroads \cite{Barnes_QRail} and 
analytically \cite{Wadati,TakaneDMPK} on the basis of the DMPK equation \cite{DMPK_D,DMPK_MPK}.
 Here we realize such an asymmetric situation in the CC model \cite{Hirose} (Fig.~\ref{fig:CCNWs_asym}).

 For asymmetric systems with two-terminal geometry, 
terminals at the ends of the system are also asymmetric in the numbers of incoming and outgoing channels
and the two-terminal conductances measured with current flowing left to right $G\_{L\to R}$, 
and right to left $G\_{R\to L}$ are related by
        \begin{equation} \label{eqn:gLR}
           G\_{L\to R}
           = G\_{R\to L} +(n\_L\^{in}-n\_L\^{out}),
        \end{equation}
where $G$ is measured in units of $e^2/h$.
 Here, ${\rm L}$ and ${\rm R}$ refer to the left and right terminals, 
and $n\_L\^{in}$ and $n\_L\^{out}$ are the number of incoming and outgoing channels, respectively, in the left terminal. 
 It follows from current conservation that 
        \begin{equation} \label{eqn:nLRio}
           n\_L\^{in}+n\_R\^{in}
           = n\_L\^{out}+n\_R\^{out}.
        \end{equation}
 Using this equation, we can rewrite eq. \eqn{gLR} as 
        \begin{equation} \label{eqn:gRL}
           G\_{R\to L}
           = G\_{L\to R} +(n\_R\^{in}-n\_R\^{out}).
        \end{equation}
 If we suppose that $n\_L\^{in} > n\_L\^{out}$, it follows that 
        \begin{equation} \label{eqn:Gminimum}
           G\_{L\to R} \ge n\_L\^{in}-n\_L\^{out}.
        \end{equation}
 Thus we expect $G\_{L\to R}$ to be finite even in the limit of infinite length (see Table~\ref{tab:Gq1D_NWs}).
 The analysis of the transmission eigenvalues 
shows that the system has $n\_L\^{in}-n\_L\^{out}$ perfectly conducting channels \cite{Barnes_QRail,TakaneDMPK,Hirose}. 
 However, the formula \eqn{Gminimum} makes it appear that 
this property is a consequence of the asymmetry of the terminals rather than the sample.

        \begin{table}[tb] 
         \caption{The two-terminal conductance in quasi-one dimensional 
                  symmetric ($n\_L\^{in}-n\_L\^{out}=0$, see Fig.~\ref{fig:CCNWs}) and
                  asymmetric ($n\_L\^{in}-n\_L\^{out}=1$, see Fig.~\ref{fig:CCNWs_asym}) CC models.}
         \begin{center} \begin{tabular}{cccc}
           \bhline
            &  $p \to 0$      &  $p=0.5$               &  $p \to 1$ \\ \hline\\[-2.6mm]
            Symmetric  
            & $G\^{q1D}=0$           & $G\^{q1D}=0$           & $G\^{q1D}=1$           \\[1.4mm]
            & $G\_{L\to R}\^{q1D}=1$ & $G\_{L\to R}\^{q1D}=1$ & $G\_{L\to R}\^{q1D}=1$ \\[0.8mm]
            \raisebox{2.7mm}[0mm][0mm]{Asymmetric}    
            & $G\_{R\to L}\^{q1D}=0$ & $G\_{R\to L}\^{q1D}=0$ & $G\_{R\to L}\^{q1D}=0$ \\[1mm]
           \bhline
         \end{tabular} \end{center} \vspace{-3mm}
        \label{tab:Gq1D_NWs}
        \end{table}

 In this paper, 
we calculate the point-contact conductance $G\_{pc}$ of an asymmetric CC network model.
 The point-contact conductance is the conductance measured between two interior probes \cite{Janssen,Klesse}.
 Just like the probes of a scanning tunneling microscope, 
the probes make contact with the sample at a point.
 The probes work as symmetric terminals ($n\_L\^{in}=n\_L\^{out}=n\_R\^{in}=n\_R\^{out}=1$) 
and $G\_{pc}$ varies between $0$ and $1$ in units of $e^2/h$. 
 In the next section, we explain how to calculate the point-contact conductance.
 In \S~\ref{sec:RES}, we show that 
the asymmetry of the network is reflected in a broad distribution of $G\_{pc}$ 
with finite averaged values in the long distance limit. 
 In the final section, we summarize and conclude.

\section{Method} \label{sec:Method}
 We denote the numbers of links in the $x$ and $y$ directions by $L_x$ and $L_y$, respectively.
 We impose 
periodic boundary condition (PBC) in the $x$ direction and 
fixed boundary conditions in the $y$ direction.
 This corresponds to a ring geometry.
 For PBC in the $x$ direction, $L_x$ must be even.
 We regard the links at $x=L_x+1$ as the ones at $x=1$.
 In the standard CC model, $L_y$ is even and the system is symmetric.
 Here we set $L_y$ odd so that the system is asymmetric.
 The state of the network is specified by the complex current amplitudes $c_i$ on the $L_x\times L_y=N$ links.

\subsection{Point-contact conductance}
\begin{figure}[tb] 
         \begin{center}\vspace{-0mm}
          \begin{tabular}{c}
            \includegraphics[width=230pt]{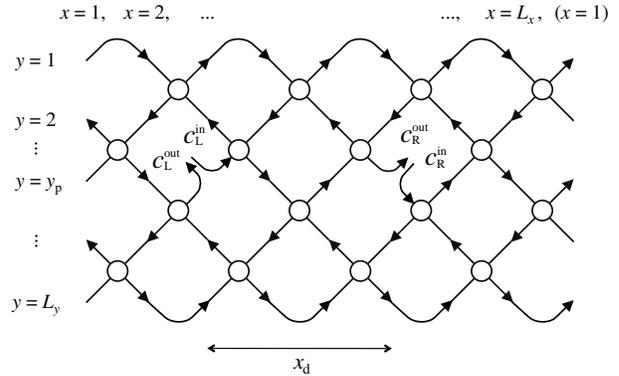}
          \end{tabular}
         \end{center}\vspace{-03mm}
        \caption{A schematic of the asymmetric CC model with point-contacts}\vspace{-2mm}
        \label{fig:CCnetPC}
        \end{figure}

 To introduce point-contacts into the network \cite{Janssen,Klesse}, 
we cut link ${\rm L}$ at $(x_1,y_1)$ 
and link ${\rm R}$ at $(x_2,y_2)$.
 We then define incoming current amplitudes $c\_L\^{in}, c\_R\^{in}$ and 
outgoing current amplitudes $c\_L\^{out}, c\_R\^{out}$ on the corresponding links (Fig.~\ref{fig:CCnetPC}).
%
 The current amplitudes satisfy the equation
        \begin{equation}\renewcommand{\arraystretch}{0.9}  \label{eqn:largeS}
         \begin{pmatrix}
          c_1 \\ c_2 \\[-1mm] \v... \\ c\_L\^{out} \\[-1mm] \v... \\ c\_R\^{out} \\[-1mm] \v... \\[-1mm] c_N
         \end{pmatrix}
         \!= {\bf S}\!
         \begin{pmatrix}
          c_1 \\ c_2 \\[-1mm] \v... \\ c\_L\^{in}  \\[-1mm] \v... \\ c\_R\^{in}  \\[-1mm] \v... \\[-1mm] c_N
         \end{pmatrix},
        \end{equation}
where ${\bf S}$ is the $N\times N$ scattering matrix consisting of $2\times 2$ scattering matrices ${\bf s}$ at each node.
 For given $(c\_L\^{in},c\_R\^{in})$, 
the remaining current amplitudes $(c_1, c_2, \c..., c\_L\^{out}, \c..., c\_R\^{out}, \c..., c_N)$ are uniquely determined 
by the following set of $N$ simultaneous linear equation with $N$ unknowns
        \begin{equation}\renewcommand{\arraystretch}{0.9}
         \begin{pmatrix}
          c_1 \\ c_2 \\[-1mm] \v... \\ c\_L\^{out} \\[-1mm] \v... \\ c\_R\^{out} \\[-1mm] \v... \\[-1mm] c_N
         \end{pmatrix}
         \!-{\bf S}\!
         \begin{pmatrix}
          c_1 \\ c_2 \\[-1mm] \v... \\ 0          \\[-1mm] \v... \\ 0          \\[-1mm] \v... \\[-1mm] c_N
         \end{pmatrix}
         \!= {\bf S}\!
         \begin{pmatrix}
          0   \\ 0   \\[-1mm] \v... \\ c\_L\^{in}  \\[-1mm] \v... \\ c\_R\^{in}  \\[-1mm] \v... \\[-1mm] 0
         \end{pmatrix}.
        \end{equation}
 As a consequence of the structure of these equations, 
there is a linear relationship between the incoming and outgoing current amplitudes
        \begin{equation}
         \begin{pmatrix}
          c\_{L}\^{out} \\
          c\_{R}\^{out}
         \end{pmatrix}
         =      
         \begin{pmatrix}
          r & t^* \\
          t & -r^*
         \end{pmatrix}
         \begin{pmatrix}
          c\_{L}\^{in} \\
          c\_{R}\^{in}
         \end{pmatrix}.
        \end{equation}
 The most straightforward way to calculate the transmission coefficient is to set
        \begin{equation} \label{eqn:supplyCurrent}
         \begin{pmatrix}
          c\_{L}\^{in} \\
          c\_{R}\^{in}
         \end{pmatrix}
         =
         \begin{pmatrix}
          1 \\
          0
         \end{pmatrix},
        \end{equation}
so that
        \begin{equation}
         t=c\_{R}\^{out}.
        \end{equation}
 The point-contact conductance $G\_{pc}$ is given by
        \begin{equation}
         G\_{pc} = |t|^2,
        \end{equation}
in units of $e^2/h$.

\section{Results} \label{sec:RES}
\subsection{Distribution of point-contact conductance}
\begin{figure}[b] 
         \begin{center}\vspace{-2mm}
          \begin{tabular}{c}
            \includegraphics[width=50mm]{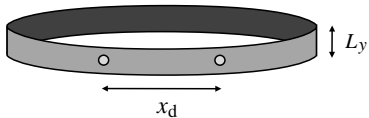}
          \end{tabular}
         \end{center}\vspace{-5mm}
        \caption{The geometry of our model. Coordinate
                 $x$ is measured along the ring and
                 $y$ across the ring.
                 The circumference of the ring is $L_x$ and the width is $L_y$. 
                 The distance between the contacts is $x\_d$ ($<L_x/2$).}\vspace{-1mm}
        \label{fig:ring}
        \end{figure}

 The point-contact conductance depends on the positions $(x_1,y_1)$ and $(x_2,y_2)$ of contacts 
in addition to the parameters of the network $L_x$, $L_y$, and $p$, 
        \begin{equation}
         G\_{pc} = G\_{pc}(x_1,y_1,x_2,y_2,L_x,L_y,p).
        \end{equation}
 This is a sample dependent quantity.
 If we average over disorder, translational symmetry is recovered, and
the averaged conductance $\braket<G\_{pc}>$ depends only on the distance 
$|x_1-x_2| \equiv x\_d$ for fixed $y_1$ and $y_2$ (see Fig.~\ref{fig:ring}).
 Taking $y_1\!=\!y_2\equiv y\_p$, 
$\braket<G\_{pc}>$ is a function of $x\_d$, $y\_p$, $L_x$, $L_y$, and $p$
        \begin{equation}
         \braket<G\_{pc}> = F(x\_d,y\_p,L_x,L_y,p).
        \end{equation}
 For convenience, we consider only two values of $y\_p$, 
corresponding to edge conductance $G\_{pc}\^e$; $y\_p=1$ and bulk conductance $G\_{pc}\^b$; $y\_p=\frac{L_y+1}{2}$.

 In the insulating limits $p\to 0$ and $p\to 1$, 
only the edge channels 
(in Fig.~\ref{fig:CCNWs}(c), Figs.~\ref{fig:CCNWs_asym}(a) and \ref{fig:CCNWs_asym}(c)) 
carry current and the point-contact conductance is bi-modal (see Table~\ref{tab:Gpc_table}).
        \begin{table}[t] 
         \vspace{-4mm}\caption{The point-contact conductances in the insulating limits 
                 (see Figs.~\ref{fig:CCNWs} and \ref{fig:CCNWs_asym}).
                  $G\_{pc}$ is unity only when the contacts are directly attached to the edge states.
                  Otherwise, $G\_{pc}=0$.}
         \begin{center} \begin{tabular}{rcc}
           \bhline
                         & $p \to 0$           &  $p \to 1$                           \\ \hline
            Symmetric  \ &       0             &  $\delta_{y\_p 1}+\delta_{y\_p L_y}$ \\
            Asymmetric \ & $\delta_{y\_p L_y}$ &  $\delta_{y\_p 1}$                   \\[0mm]
           \bhline
         \end{tabular} \end{center}\vspace{-5mm}
        \label{tab:Gpc_table}
        \end{table}

\begin{figure}[b] 
         \vspace{-2mm}\begin{center}
          \begin{tabular}{r}
            \includegraphics[width=70mm]{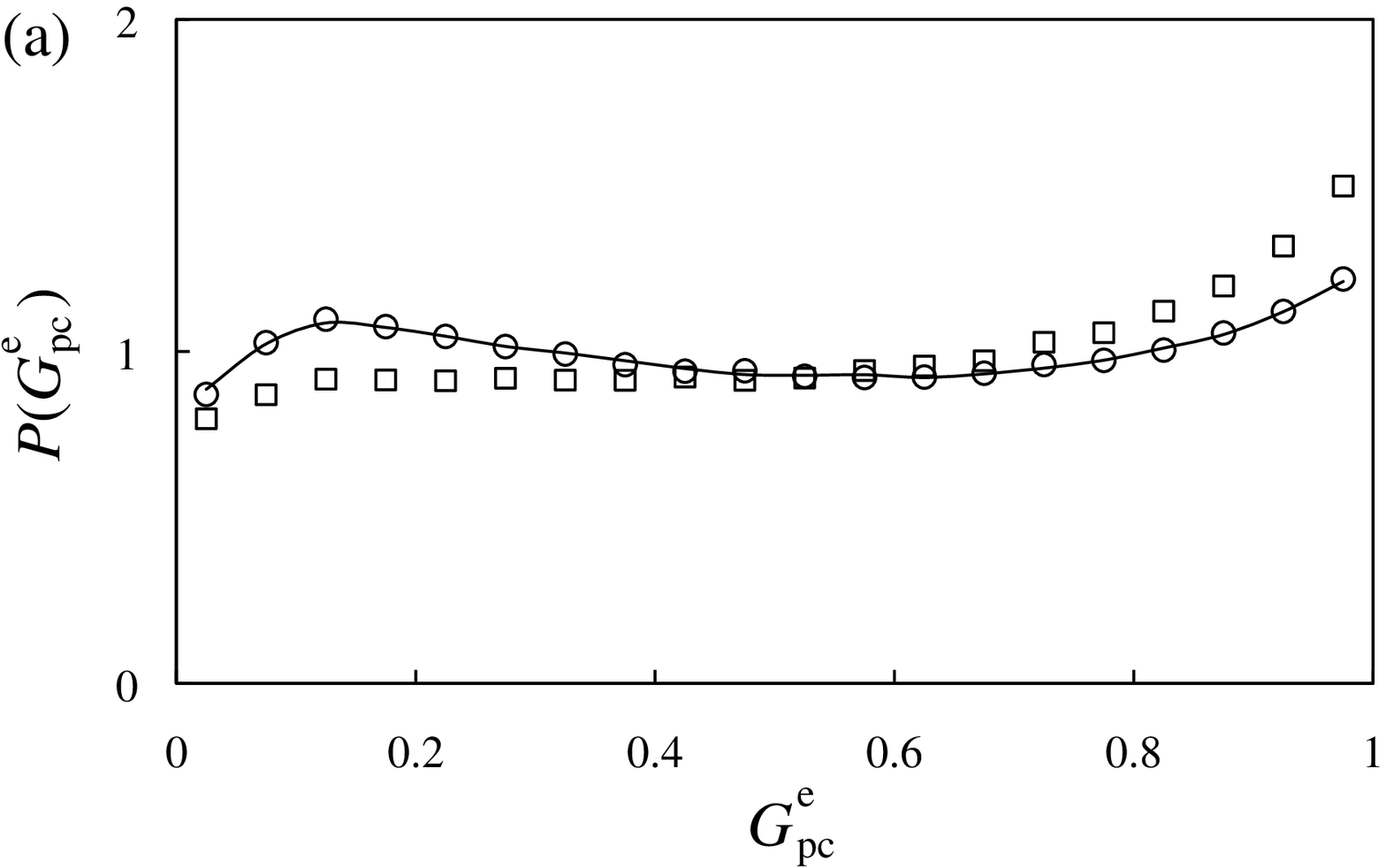} \\[2mm]
            \includegraphics[width=70mm]{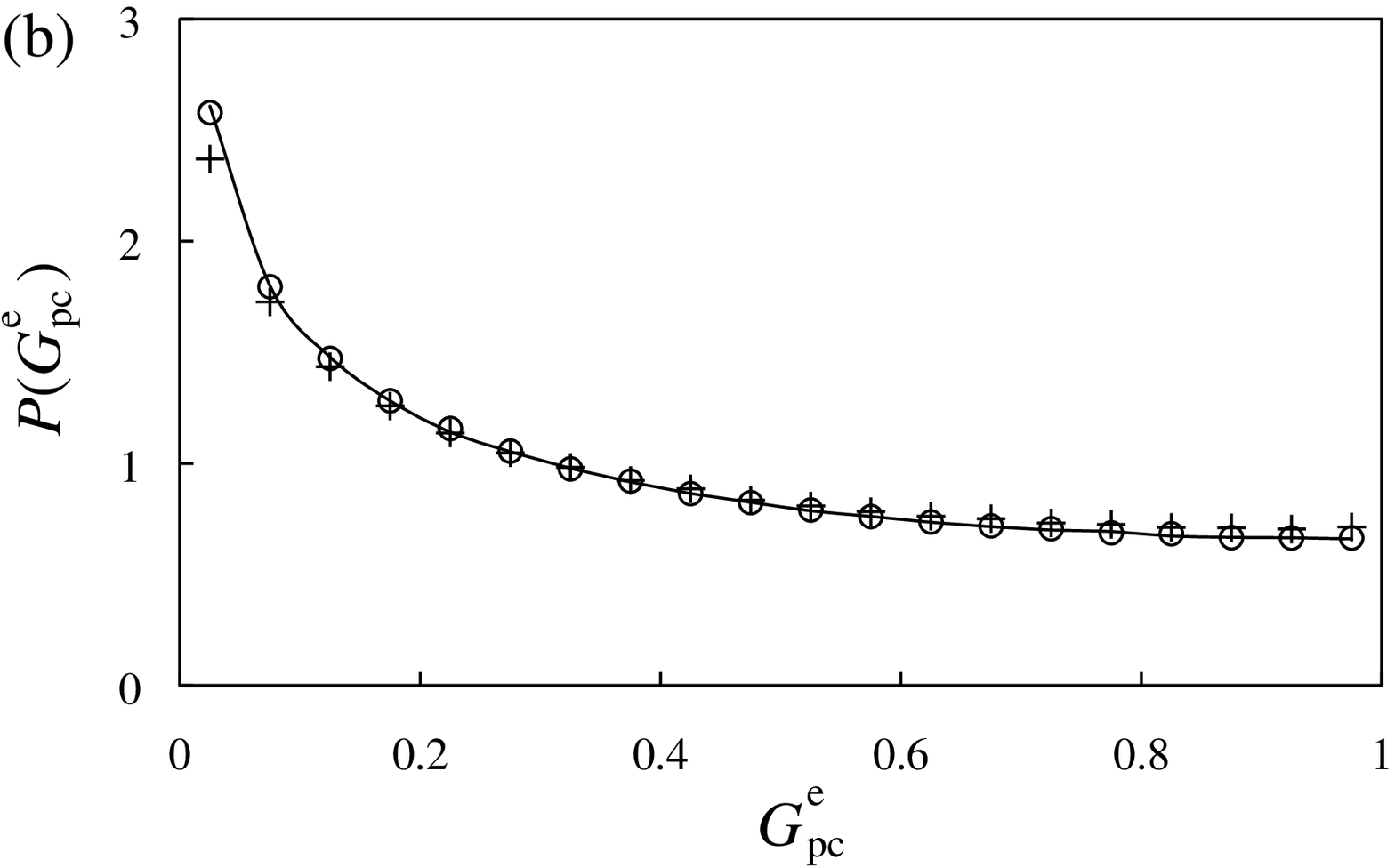}
          \end{tabular}
         \end{center}\vspace{-3mm}
        \caption{The distribution of the edge conductance 
                 for $L_y=9$, (a) $x\_d=3$ for $L_x=$ $12$({\tiny $\square$}), $36$($\circ$), $900$(---), 
                 and (b) $x\_d=9$ for $L_x=$ $24$({\tiny $+$}), $36$({\tiny $\times$}), $900$(---).
                 Ensemble averages over 1,000,000 systems have been taken.
                 Irrespective of the value of $x\_d$, 
                 the dependence on $L_x$ disappears for $L_x\gg L_y$.}\vspace{-4mm}
        \label{fig:PG_Lx-dep}
        \end{figure}

\begin{figure}[t] 
         \begin{center}
          \begin{tabular}{r}
            \includegraphics[width=78mm]{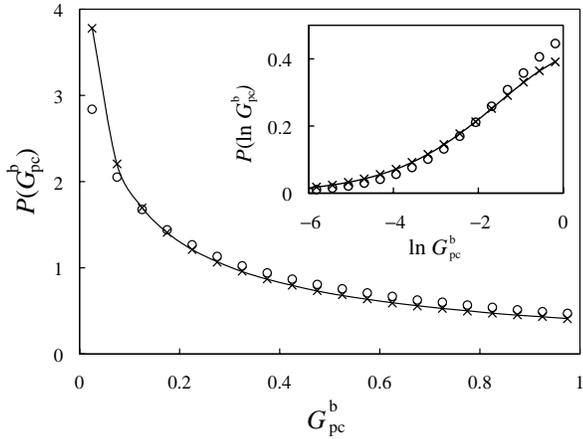}
          \end{tabular}
         \end{center}\vspace{-4mm}
        \caption{The distribution of the bulk conductance 
                 for $L_y=9$, for $x\_d=$ $9$({\tiny $\circ$}), $45$({\tiny $\times$}), $450$(---).
                 Ensemble averages over 5,000,000 systems have been taken.
                 Inset: Distribution of logarithms of point-contact conductance.
                 For large $x\_d$, we see convergence to a broad distribution.
                 }\vspace{-3mm}
        \label{fig:PG_xd-dep}
        \end{figure}

 At criticality $p=0.5$, the form of the point-contact conductance distribution $P(G\_{pc})$ is more complicated.
 The distributions of the edge conductance 
obtained from numerical simulations of systems with $L_y=9$ and various $x\_d$ and $L_x$ 
are shown in Fig.~\ref{fig:PG_Lx-dep}.
 Similar results are obtained for the distribution 
in the bulk (see Fig.~\ref{fig:PG_xd-dep}).
 For $L_x\gg L_y$, the distribution tends to a limiting form that depends on 
$x\_d$, $y\_p$, and $L_y$.
 For $x\_d \gg L_y$, the $x\_d$ dependence of this limiting distribution disappears (Fig.~\ref{fig:PG_xd-dep}).
 Surprisingly, there is no self-averaging of the point-contact conductance even when $x\_d\rightarrow \infty$ and the distribution remains broad.


\begin{figure}[t] 
         \begin{center}
          \begin{tabular}{c}
            \includegraphics[width=74mm]{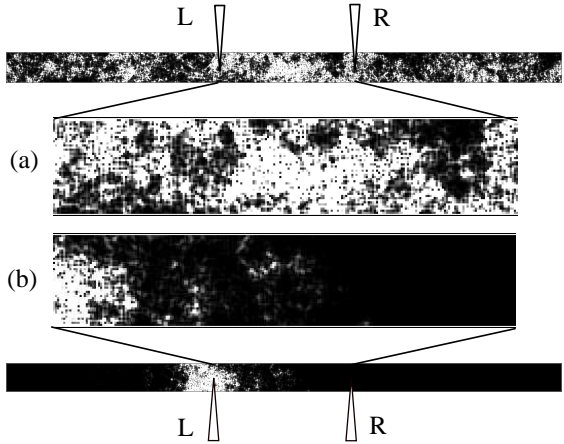}
          \end{tabular}
         \end{center}\vspace{-4mm}
        \caption{The squared flux amplitude $|c_i|^2$ at criticality in 
                 (a) the asymmetric CC network of $(L_x,L_y)=(900,45)$ and 
                 (b) the symmetric CC network of $(L_x,L_y)=(900,44)$.
                 Darker areas correspond to lower squared amplitudes.
                 Probes, indicated by wedges, are attached at the middle of the system.
                 Current flows from the left probe to the right probe ($c\_{L}\^{in}=1$, $c\_{R}\^{in}=0$, see eq. \eqn{supplyCurrent}).
                 }\vspace{-3mm}
        \label{fig:snap}
        \end{figure}

 In Fig.~\ref{fig:snap}, we show the squared flux amplitudes $|c_i|^2\,(i=1,...,{\rm L}\^{out},...,{\rm R}\^{out},...,N)$. 
 Note that in the asymmetric case, the current is distributed all across the sample even in the limit $x\_d \gg L_y$ (Fig.~\ref{fig:snap}(a)).
 This is in sharp contrast to the symmetric case where the current quickly decays (Fig.~\ref{fig:snap}(b)). 

\subsection{Dependence of $\braket<G\_{pc}>$ on $x\_d$} 
\begin{figure}[t] 
         \begin{center}
          \begin{tabular}{c}
            \includegraphics[width=70mm]{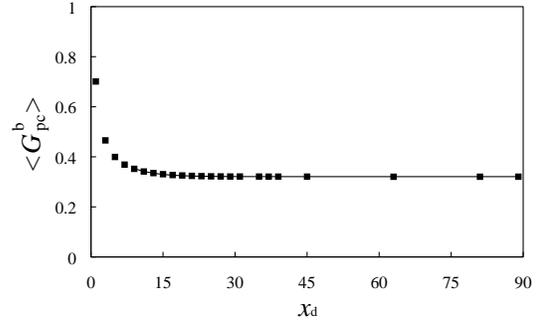}
          \end{tabular}
         \end{center}\vspace{-2mm}
        \caption{
         The average of the bulk conductance as a function of $x\_d$ for $L_y=9$.
         The data are an average over an ensemble of 2,000,000 systems.
         The solid line is the fit to eq. \eqn{fit1} 
         with $\braket<G_\infty\^b> = 0.3208\pm 0.0001$, $b=0.670\pm 0.017$, and $\lambda=4.67\pm 0.05$ (goodness of fit $Q=0.81$).
        }\vspace{-3mm}
        \label{fig:Gpc_x-dep_L9}
        \end{figure}

 To quantify how the conductance distribution converges to its limiting form, 
we study the $x\_d$ dependence of the averaged conductance.
 We have found that the averaged conductance converges exponentially, 
        \begin{equation} \label{eqn:fit1}
         \braket<G\_{pc}> = \braket<G_\infty> \left[ 1+b\exp\(-\frac{x\_d}{\lambda}\) \right] .
        \end{equation}
 An example is shown in Fig.~\ref{fig:Gpc_x-dep_L9}.
 Note that the values of $\braket<G_\infty>$, $b$, and $\lambda$ depend, in principle, on $y\_p$ and $L_y$.
 We emphasize that $G_\infty=0$ in the symmetric CC model.

\subsection{Dependence of $\braket<G\_{pc}>$ on $L_y$}
 We now analyze the $x\_d$-dependence of averaged edge conductance $\braket<G\_{pc}\^e>$
and similarly for the bulk conductance $\braket<G\_{pc}\^b>$ for various $L_y$.

\begin{figure}[b] 
         \begin{center}
          \begin{tabular}{r}
            \includegraphics[width=75mm]{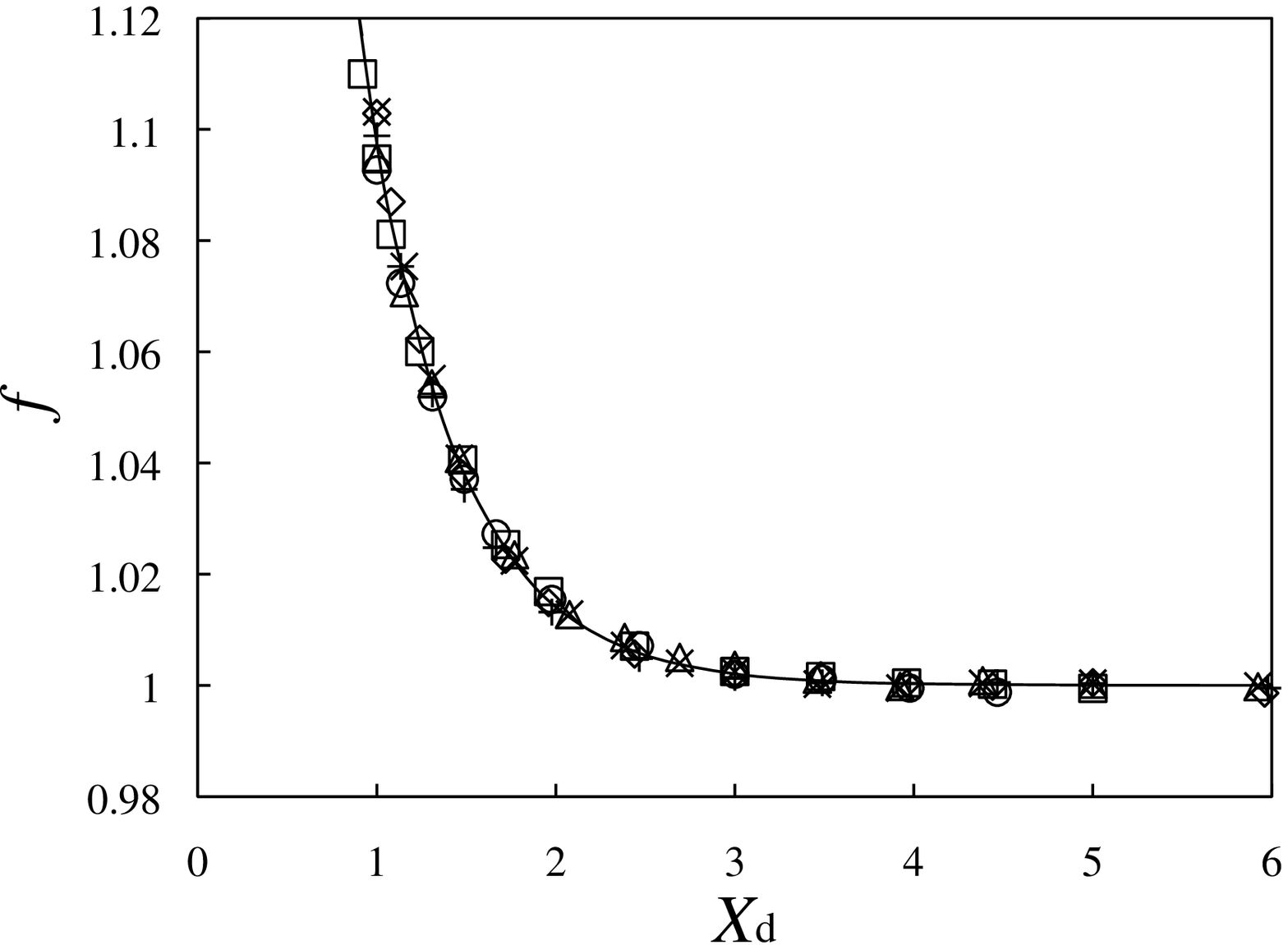}
          \end{tabular}
         \end{center}\vspace{-2mm}
        \caption{The ratio
        $f=\braket<G\_{pc}>/\braket<G_\infty>$ as a function of $X\_d=x\_d/L_y$ for 
        $L_y=$ $13$($\times$), $25$($\diamond$), $45$($+$) (edge conductance)
        and $L_y=$ $13$({\tiny $\triangle$}), $25$({\tiny $\square$}), $45$($\circ$)  (bulk conductance).
        The solid line is a fit to eq. \eqn{f_para-form} (goodness of fit probability $Q = 0.59$).}\vspace{-2mm}
        \label{fig:x-dep_norm}
        \end{figure}

 Scaling form describing the dependence of $\braket<G\_{pc}>$ on $x\_d$ and $L_y$ can be derived by assuming following factorization,
        \begin{eqnarray}
         \braket<G\_{pc}>=h(y\_p,L_y)\,f(x\_d,L_y).
        \end{eqnarray}
 To eliminate the ambiguity in this factorization, we set $f(x\_d\to \infty,L_y)=1$.
 Taking the limit $x\_d \to \infty$,  
        \begin{eqnarray}
         \braket<G\_{\infty}> = h(y\_p,L_y).
        \end{eqnarray}
 Comparing with eq. \eqn{fit1}, we can write 
        \begin{eqnarray}\label{eqn:f_para-form}
         &f(x\_d,L_y) = 1 + b\exp\!\(-X\_d/\Lambda\)\!,\ \\
         &X\_d = x\_d/L_y,\quad      \Lambda = \lambda/L_y.
        \end{eqnarray}
 We have found that data for different $x\_d$ and different $L_y$ 
collapse onto a single curve (Fig.~\ref{fig:x-dep_norm}) with the following values,
        \begin{eqnarray} \label{eqn:f_para-param}
          b = 0.675\!\pm\! 0.019,\ \Lambda = 0.518\!\pm\! 0.006.
        \end{eqnarray}
%

\subsection{Dependence of $\braket<G_\infty>$ on $L_y$}  
\begin{figure}[t] 
         \begin{center}
          \begin{tabular}{c}
            \includegraphics[width=78mm]{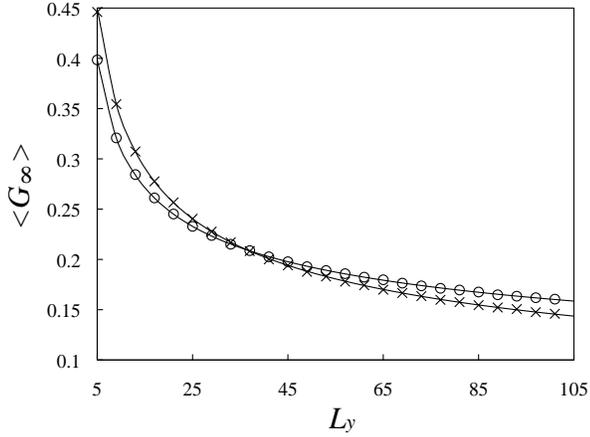}
          \end{tabular}
         \end{center}\vspace{-4mm}
        \caption{The imiting value of the point-contact conductance $\braket<G_\infty>$ 
                 for edge($\times$) and bulk($\circ$) as a function of $L_y$.
                 The data are estimated from disorder averages over 400,000 systems.
                 Solid lines are the fits to eq. \eqn{Ginf_Ly-dep}.}\vspace{-5mm}
        \label{fig:Ginf_Ly-dep}
        \end{figure}

 The dependence of $\braket<G_\infty>$ on $L_y$ for edge and bulk is shown in Fig.~\ref{fig:Ginf_Ly-dep}.
 We have found that the following form 
        \begin{eqnarray}\label{eqn:Ginf_Ly-dep}
         \braket<G_\infty^{i}>    = C_0^{i}    L_y{}^{\alpha^{i}}    + \frac{C_1^{i}}{L_y}
        \end{eqnarray}
fits our data.
 The best-fit values of parameters are listed in Table~\ref{tab:fit_inf_param}. 
 Here $i$ denotes whether $G\_{pc}\^e$ (edge) or $G\_{pc}\^b$ (bulk).
 The first term is a non-trivial power law decay 
that reflects the multi-fractal \cite{aoki83,Evers08,Evers_review,Obuse_MFCFT,Obuse_CCMF}
nature of the conducting states.
 The second term is a correction for the discreteness of the model and the effect of the boundary \cite{Slevin00}. 
 The difference between edge and bulk conductance may originate from 
the difference between the surface and bulk multi-fractality \cite{Obuse_CCMF}. 

        \begin{table}[t] 
         \caption{Best-fit parameters for eq. \eqn{Ginf_Ly-dep}.}
         \begin{center}
         \begin{tabular}{@{\ }c@{\ \ \,}c@{\ \ \,}c@{\ \ \,}c@{\ \ \,}c@{\ }}  \bhline
                & $C_0$               & $\alpha$           & $C_1$               & $Q$\\  \hline
          Edge  & $0.6907 \!\pm\! 0.0053$ & $-0.3413 \!\pm\! 0.0016$ & $0.2531 \!\pm\! 0.0126$ & $0.59$ \\
          Bulk  & $0.4590 \!\pm\! 0.0018$ & $-0.2338 \!\pm\! 0.0009$ & $0.4167 \!\pm\! 0.0047$ & $0.70$ \\  \bhline
         \end{tabular}
         \end{center}\vspace{-4mm}
        \label{tab:fit_inf_param}
        \end{table}
%

\section{Summary and Concluding Remarks}  \label{sec:SUM}
 In this paper, 
we have calculated the point-contact conductance $G\_{pc}$ in the asymmetric Chalker-Coddington network model 
and found a novel metallic behavior. 
 In contrast to the symmetric Chalker-Coddington network model, 
the point-contact conductance distribution converges to a broad distribution for a large separation of the contacts.
 This is true both for contacts attached to the bulk and the edges of the sample at criticality.
 This broad distribution reflects the nature of the current distribution of the perfectly conducting state.
 We have also studied the averaged point-contact conductance 
in the limit of a large circumference, and found a scaling form.
 Both $\braket<G\^e_\infty>$ and $\braket<G\^b_\infty>$ show 
non-trivial power law decay eq.~\eqn{Ginf_Ly-dep} (see also Fig.~\ref{fig:Ginf_Ly-dep}), 
which is a characteristic of criticality.

 So far we have focused on criticality $p=0.5$.
 When $p$ deviates from $0.5$, 
the states are localized in the transverse direction. 
 When system width $L_y$ exceeds the transverse localization length, 
the perfectly conducting state is localized along one of the edges, 
$y=1$ for $p>0.5$ and $y=L_y$ for $p<0.5$ (Figs.~\ref{fig:CCNWs_asym}(a) and \ref{fig:CCNWs_asym}(c) are extreme examples). 
 In this case, $\braket<G\_{pc}>$ decays quickly with the distance from the conducting edge.
 Broad distributions $P(G\_{pc})$ are observed only 
when we attach contacts near the conducting edge.
 Note that even in the ordinary quantum Hall effect, 
such fluctuations in point-contact conductances are expected near the edges.

 It is known that the conductance distribution is sensitive to the symmetry class (unitary, orthogonal, or symplectic) 
classified according to the presence or absence of time-reversal and spin-rotation symmetries.
 Since the time-reversal symmetry is broken in the scattering matrix eq.~\eqn{largeS}, 
the asymmetric Chalker-Coddington model 
belongs to the unitary class \cite{Takane_3eCC}.
 A perfectly conducting channel also arises in the symplectic class, 
which is realized in carbon nanotubes \cite{ando98,AndoSuzuura_PCC,suzuura02,takane04}. 
 The distribution of the point-contact conductance in the symplectic class, 
especially in the metal phase, 
may also be worth investigating.

\section*{Acknowledgment}
 This work was supported by Grant-in-Aid No. 18540382.
 We would like to thank Dr. H. Obuse and Mr. K. Hirose for useful discussions and fruitful comments.


\label{lastpage}
\end{document}